\documentclass[aps,groupedaddres8s,fleqn,nofootinbib,twocolumn]{revtex4}
\usepackage{graphicx}
\usepackage{xcolor}
\usepackage{amsmath,amssymb}
\usepackage{float}
\usepackage{amsfonts}
\usepackage{verbatim}
\usepackage{flushend}
\usepackage{balance}
\usepackage{endnotes}
\usepackage{footnote}
\usepackage{adjustbox}
\usepackage{gensymb}
\usepackage{float}
\usepackage{epigraph}
\usepackage{xcolor}
\usepackage[utf8]{inputenc}
\usepackage{setspace}

\begin{document}
\title{Extrinsic and intrinsic effects setting viscosity in complex fluids and life processes: the role of fundamental physical constants}
\author{K. Trachenko$^1$}
\author{P. G. Tello$^2$}
\author{S. A. Kauffman$^3$}
\author{S. Succi$^{4,5}$}
\address{$^1$ School of Physical and Chemical Sciences, Queen Mary University of London, Mile End Road, London, E1 4NS, UK}
\address{$^2$ CERN EU Office, Geneva, Switzerland}
\address{$^3$ Biochemistry and Biophysics (Emeritus), University of Pennsylvania, USA}
\address{$^4$ Center for Life Nano Science@La Sapienza, Istituto Italiano di Tecnologia, Viale Regina Elena 291, Rome, 00161, Italy}
\address{$^5$ Department of Physics, Harvard University, 17 Oxford Street, Cambridge, 02138, MA, USA}

\begin{abstract}
Understanding the values and origin of fundamental physical constants, one of the grandest challenges in modern science, has been discussed in particle physics, astronomy and cosmology. More recently, it was realised that fundamental constants have a bio-friendly window set by life processes involving motion and flow. This window is related to intrinsic fluid properties such as energy and length scales in condensed matter set by fundamental constants. Here, we discuss important extrinsic factors governing the viscosity of complex fluids operating in life processes due to collective effects. We show that both extrinsic and intrinsic factors affecting viscosity need to be taken into account when estimating the bio-friendly range of fundamental constants from life processes, and our discussion provides a straightforward recipe for doing this. We also find that the relative role of extrinsic and intrinsic factors depends on the range of variability of these intrinsic and extrinsic factors. Remarkably, the viscosity of a complex fluid such as blood with significant extrinsic effects is not far from the intrinsic viscosity calculated using the fundamental constants only, and we discuss the reason for this in terms of dynamics of contact points between cells.
\end{abstract}

\maketitle

\section{Introduction}

Fundamental physical theories of matter and fields have about 20 fundamental physical constants such as the Planck constant $\hbar$, the electron mass $m_e$ and charge $e$ and other parameters
(see, e.g., Refs. \cite{uzanreview,uzan1,codata,ashcroft}). These constants are consistent with the observed properties of the Universe
\cite{barrow,barrow1,carrbook,finebook,carr1,carr,cahnreview,hoganreview,adamsreview,uzanreview,uzan1}.
Understanding fundamental constants, their values and origin, is considered as one of the grandest questions in science \cite{grandest}. As observed some time ago, we do not know what sort of theories we
need to explain the values and origin of fundamental constants \cite{weinberg}. The values of fundamental constants are currently considered arbitrary \cite{cahnreview}.

The role of fundamental constants was discussed in high-energy processes, including inflation and cosmology, particle physics, nuclear reactions and nuclear synthesis in stars, including, for example, the Hoyle resonance. This nuclear synthesis produces carbon, oxygen and other elements which can form molecular structures essential to life at later stages of cosmic evolution. It was observed that small relative changes of some fundamental constants would disable the essential high-energy processes above. In other words, fundamental constants are thought to be finely tuned to result in observed phenomena \cite{barrow,carrbook,finebook}.

More recently, it was proposed that condensed matter physics and liquid physics in particular gives a new insight into the fundamental constants based on life processes \cite{biofriendly}. We observe these life processes and can therefore discuss factors that enable them. Life processes need motion and flow, and dynamic viscosity $\eta$, is the central property setting this flow. The minimal viscosity was previously shown to be set by fundamental physical constants \cite{sciadv1,pt2021}. If this minimum is increased, liquid viscosity becomes larger at all conditions of pressure and temperature. Its value corresponding to disabling a life process, $\eta_d$, then puts a limit on bio-friendly values of fundamental constants. A detailed consideration of this process implies that there is a bio-friendly {\it window} where fundamental constants can vary to enable life processes in and between living cells \cite{biofriendly}.

Several effects related to viscosity such as diffusion and flow can affect a life process. For example, large viscosity corresponds to the critical value of the P\'eclet number at which the

There are several viscosity effects involved in affecting and potentially disabling a life process. For example, large viscosity corresponding to the critical P\'eclet number results in transitions related to the explosive increase of the coagulation rate in protein solutions, blood and other biological fluids \cite{zaccone-peclet}. Another example is chemical reactions: chemical reaction rates $k$ of important biological processes involving proteins and enzymes depend on $\eta$ as $k\propto\frac{1}{\eta^n}$, where $n$ can vary between 2.4 and 0.3 \cite{chemrate1} for different reactions \cite{chemrate3}. Hence, larger $\eta$ has different effects on different chemical reactions, changing the balance between reaction products and operation of those products. Depending on how important this change is, a new functioning balance may or may not be found.

The large increase of the coagulation rate and chemical reactions are two specific examples of extrinsic effects related to fluid flow. More generally, both equilibrium and transport phenomena in soft condensed matter are governed by effective equations of state and transport coefficients, reflecting collective interactions with the environment. This comes in addition to intrinsic molecular values and involves strong nonlinearities \cite{sauroOUP18,sauroOUP22,sauroEOS}. An important question is then how intrinsic and extrinsic effects compare under different conditions and how they can be combined to understand (a) viscosity at the fundamental level and (b) implications for fundamental constants.

In this paper, we discuss these extrinsic effects and find that both extrinsic and intrinsic factors affecting viscosity need to be taken into account when estimating the bio-friendly range of fundamental constants from life processes. Our discussion provides a straightforward recipe for doing this. We also find that the relative role of extrinsic and intrinsic factors depends on the range in which these intrinsic and extrinsic factors vary. Remarkably, viscosity of a complex fluid such as blood with significant extrinsic effects is not far from the intrinsic viscosity calculated using the fundamental constants only, and we discuss the reason for this in terms of dynamics of contact points between cells.

\section{Intrinsic effects}
\label{intrinsic}

We briefly recall the intrinsic effects involved in setting the minimal viscosity of fluids \cite{sciadv1,pt2021,biofriendly}. The basis for discussing this minimal viscosity has been theoretical. This is a recent result, which may come across as surprising in view of long history of research into viscosity and, more generally, theory of liquids. For this reason, we briefly expand on this point and recall the fundamental nature of problems involved in liquid theory.

In gases, interactions are weak. In solids, particle displacements are assumed to be small. In liquids, neither of these properties apply. Due to this absence of simplifying features of liquid theory (the absence of a small parameter), it was considered that a general theory of basic liquid thermodynamic properties such as energy is not feasible \cite{landaustat,pitaevskii,ahiezer}. This is in contrast to gases and solids where calculated thermodynamic properties and their temperature dependence are generally-applicable and form the basis of solid state and gas theories.

In an interacting systems, excitations or quasiparticles govern important system properties \cite{landaustat,landaustat1}. In solids, these excitations are commonly considered phonons. In liquids, phonons and their properties remained unknown for a long time. Interestingly, Sommerfeld \cite{somm} and Brillouin \cite{br1,br2,br3,brillouin,brilprb} proposed that energy and other thermodynamic properties of liquids are governed by phonons as they are in solids and sought to apply a Debye theory to liquids. This was around the same time when the basis for the modern solid state theory was set \cite{debyepaper,einstein}. The nature and operation of phonons in liquids was not clear at the time and turned out to be a formidable problem that continues to be actively researched today \cite{mybook}.

Understanding phonons in liquids involved inputs from experiments, theory and modelling. An important insight from this research is that the phase space taken by these phonons is not fixed as in solids but is {\it variable} \cite{ropp,mybook,proctor1,proctor2,chen-review}. In particular, the phase space reduces with temperature, and this has a general implication for most important liquid properties. For example, the calculated specific heat of classical liquids universally decreases with temperature, in quantitative parameter-free agreement with a wide range of experimental data \cite{ropp,mybook,proctor1,proctor2}.

This recent new understanding of liquids has brought about the concept of the minimal quantum viscosity and its relation to fundamental physical constants \cite{sciadv1,pt2021,biofriendly}. The minimal kinematic viscosity, $\nu_{min}$, is set by two parameters characterising a condensed matter phase: the interatomic separation $a$ and the Debye vibration frequency $\omega_{\rm D}$ as:

\begin{equation}
\nu_{min}=\frac{1}{2\pi}\omega_{\rm D}a^2
\label{num}
\end{equation}

Relating $a$ to the Bohr radius

\begin{equation}
a_{\rm B}=\frac{4\pi\epsilon_0\hbar^2}{m_e e^2}
\label{bohr}
\end{equation}

\noindent where $e$ and $m_e$ are electron charge and mass, and $\omega_{\rm D}$ to the characteristic cohesive energy set by the Rydberg energy

\begin{equation}
E_{\rm R}=\frac{m_ee^4}{32\pi^2\epsilon_0^2\hbar^2}
\label{rydberg}
\end{equation}

\noindent gives

\begin{equation}
\nu_{min} = \frac{1}{4 \pi} \frac{\hbar}{\sqrt{m_e m}}
\label{num1}
\end{equation}

\noindent where $m$ is the molecule mass \cite{sciadv1}.

Noting that the scale of $m$ is set by the proton mass $m_p$, Eq. \eqref{num1} gives rise to fundamental kinematic viscosity $\nu_f$:

\begin{equation}
\nu_{min} = \frac{1}{4 \pi} \frac{\hbar}{\sqrt{m_e m_p}}
\label{num2}
\end{equation}

\noindent of about $10^{-7}$ m$^2$/s \cite{sciadv1}.

$\nu$ and its minimal value in Eq. (\ref{num1}) govern the time-dependent non-equilibrium flow. $\nu$ also sets the Reynolds number and Kolmogorov scale of turbulence. The steady flow is set by the dynamic viscosity $\eta$. The minimum of $\eta$, $\eta_{m}$, can be evaluated as $\eta_m=\nu_m\rho$, where $\rho$ is density $\rho\approx\frac{m}{a_{\rm B}^3}$. Assuming $m=Am_p$, where $A$ is atomic number and setting $A=1$ for the purpose of the following discussion, this gives

\begin{equation}
\eta_{min}\propto\frac{e^6}{\hbar^5}\sqrt{m_pm_e^5}
\label{etamin}
\end{equation}

Another useful property is the diffusion constant $D$. Using Eq. (\ref{etamin}) and the Stokes-Einstein relation $D=\frac{k_{\rm B}T}{6\pi r\eta}$, where $r$ is the radius of a moving particle, gives

\begin{equation}
D_{max}\propto\frac{1}{\eta_{m}}\propto\frac{\hbar^5}{e^6}\frac{1}{\sqrt{m_pm_e^5}}
\label{dmax}
\end{equation}

We see that the intrinsic effects affecting liquid flow, the minimal values of kinematic and dynamic viscosity and the maximal diffusion constant are related to the length and energy scales involved in chemical bonds in the liquid. These scales are relatable to fundamental constants, and so are $\nu_{min}$, $\eta_{min}$ and $D_{max}$.

\section{Extrinsic effects: blood flow as a case study}

The extrinsic effects affecting liquid flow include those operating in complex fluids such as, for example, blood.
Blood flow delivers nutrients in any organism and is therefore related to the metabolism, the essence
of life \cite{lanebook}, along with genetics. Blood is not a simple fluid with a given viscosity, but a dense suspension of red cells, platelets and plasma particles \cite{sauroBLOOD,bloodvisc,noirezblood1,noirezblood2} whose {\it effective} viscosity is affected by collective many-body effects.
As a result, blood is a non-newtonian fluid, which does not deform in linear proportion to the stress acting upon it. Stated differently, the effective viscosity is a property of the {\it flow process}, not of the fluid substance.

We briefly recall some of the basic extrinsic properties. A Newtonian fluid under the Couette flow (flow between two oppositely moving flat plates) obeys the following relation:

\begin{equation}
\label{NEWTON}
F_x = \mu_0 A \frac{\partial u_x}{\partial y}
\end{equation}

\noindent where $F_x$ is the force along the mainstream direction $x$, $A$ is the area, $u_x$ is the flow speed, $\mu_0 = \rho \nu_0$ is the bare dynamic viscosity and $\rho$ is density. For a simple fluid, $\mu_0$ is a numerical coefficient (at a given temperature), whose value can be traced to the quantum diffusivity $D_q \equiv \hbar/m$.

The volume fraction of red cells in the blood (hematocrit) is about $\phi \sim 0.45$, which qualifies blood as a dense suspension in which the average gap between two cells is much smaller than their diameter. Consider a sphere of diameter $2R$ located in the center of a cubic box of side $L$. The volume fraction of the sphere in this simple-cubic configuration is $\phi = \frac{\pi}{6} \frac{(2R)^3}{L^3}$. The gap between two spheres is $h=L-2R$, hence $\frac{h}{2R}=\frac{L}{2R}-1=\left(\frac{\pi}{6\phi}\right)^{\frac{1}{3}}-1$. With $\phi \sim 0.45$, this gives $\frac{h}{2R}\sim 0.05$, implying that cells are
constantly in near-touch 
so that their motion is
strongly affected by many-body effects.

In particular, at low shear rate say $S =0.1$ 1/s, red cells tend to aggregate, forming clusters which withstand fluidity. Upon increasing the shear rate, clusters break up and blood flows more easily: blood is a so-called shear-thinning fluid.
This means that viscosity no longer depends on intrinsic effects and fundamental constants only, but acquires an additional nonlinear dependence on shear $S=\partial_y u_x$.
This nonlinearity is typically expressed by a power-law relation of the form:

\begin{equation}
\eta(S) = \frac{\eta_0}{(1+S/S_0)^{\alpha}}
\label{etaext}
\end{equation}

\noindent where $S_0$ is the threshold above which the nonlinear behavior is exposed and $\alpha$ is a characteristic exponent (typically around $1/3$) and $\eta_0$ is related to the intrinsic viscosity corresponding to no shear effects.

As an example of the magnitude of this effect, an excursion of three decades in the value of $S$ leads a factor of ten change in the effective viscosity. The same liquid showing a viscosity of $60$ cP at $S=0.1$ (1/s), can lower its viscosity down to $6$ cP at $S=200$ (1/s). In Figure 1, we show the best fit of actual physiological data from reference \cite{sauroBLOOD} using expression (\ref{etaext}) with the following numerical values: $\eta_0=75$ cP, $S_0=0.1$ and $\alpha=0.35$. The fit reproduces the experimental data quite accurately.

\begin{figure}
\begin{center}
{\scalebox{0.35}{\includegraphics{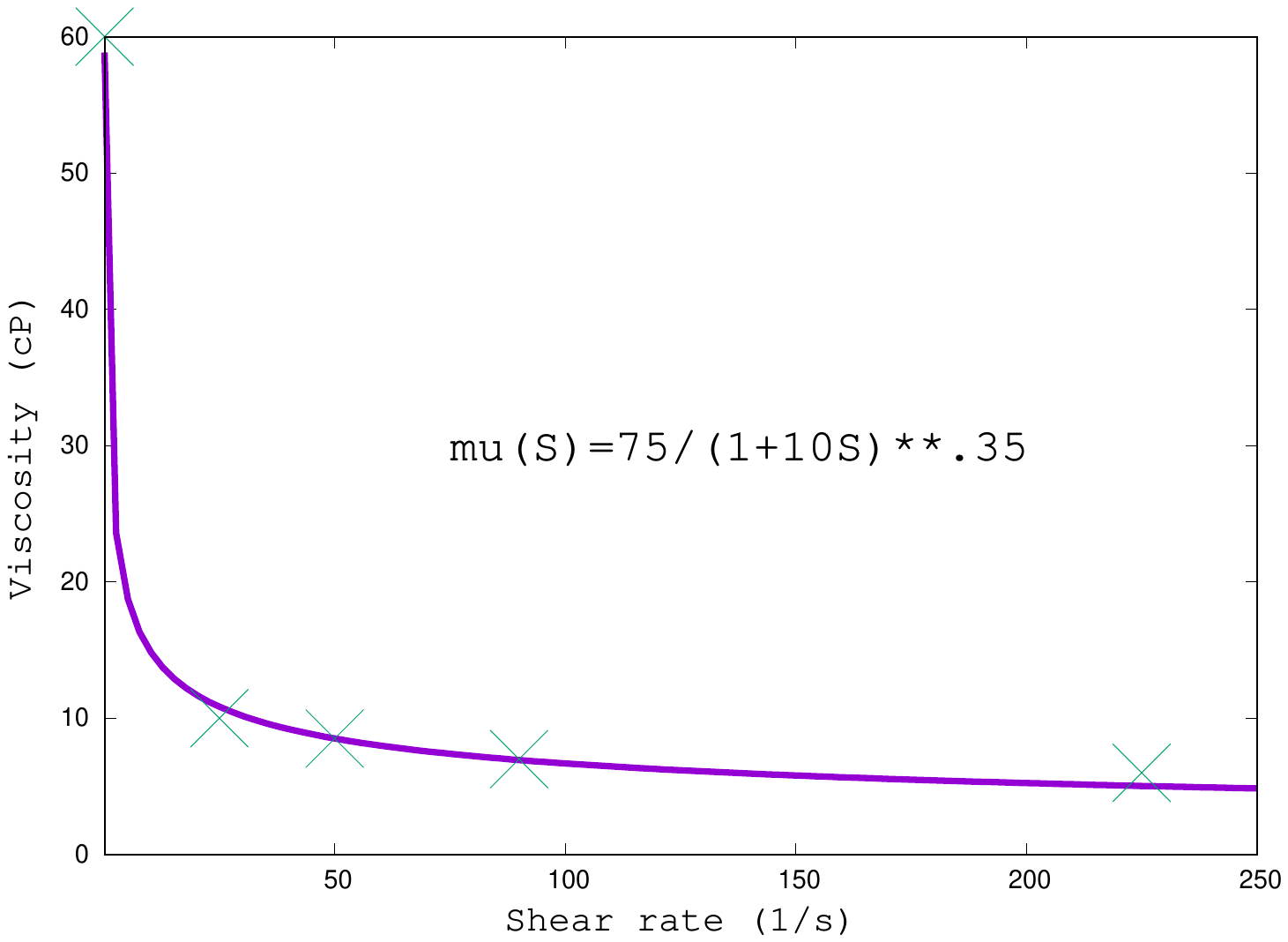}}}
\end{center}
\caption{Best fit to blood viscosity using expression \eqref{etaext}, with $\eta_0=75$ cP, $S_0=0.1$ and $\alpha=0.35$.}
\end{figure}

In addition to the above example, collective effects in dense suspensions can lead to a host of complex rheological behaviours, including yield-stress (no flow below a minimum stress threshold), nonlocal effects in space and time, hysteresis, jamming and many others, including crucial effects of deformability for many metabolic functions.
For example, local accumulation of red blood cells due to physiological imperfections can drive an untamed rise of viscosity, a situation known as clogging or jamming as mentioned in the Introduction.
In this case, the effective viscosity obeys a similar law:

\begin{equation}
\eta = \frac{\eta_0}{(1-\rho/\rho_c)^{\beta}}
\label{JAM}
\end{equation}

\noindent where $\rho_c$ is the critical density at which the viscosity formally diverges, $\beta$ is the corresponding critical exponent and $\eta_0$ is related to the intrinsic viscosity in the absence of jamming effects.

\section{Intra-cellular and anomalous diffusion}

Similar effects operate for the diffusion involving cellular diffusion. Here, the direct analogue of the Newtonian fluid is the Fick's law, which is a statement of linearity between the density gradient and the resulting mass flux $J_x$:

\begin{equation}
\label{FICK}
J_x = -D_0 \frac{\partial \rho}{\partial x}
\end{equation}

Fick's law holds as long as the gradients are sufficiently small to silence nonlinear effects and density is sufficiently below the jamming transition.

The cell is a complex and crowded environment, in which molecular motion can hardly abide by the simplicity of Fick's law \cite{sauroCROWD}. Molecules crawl in the cytoplasm, collide with obstacles and get eventually trapped in metastable states until they are released again. Mechanisms of diffusion and transport in cells include molecular motors \cite{sauroHUNT}. In these diffusive processes, the bare diffusivity $D_0$ is replaced by a density-dependent effective diffusivity $D=D(\rho)$, typically a strongly nonlinear function, eventually vanishing at a critical value, $D(\rho \to \rho_c) \to 0$ (jamming), consistent with the Stokes law:

\begin{equation}
D = \frac{k_BT}{6 \pi R \eta}
\end{equation}

\noindent where $R$ is the radius of the sphere floating in a solvent of viscosity $\eta$.

Collective effects can also extend beyond density-dependence of the diffusion coefficient and alter the nature of the diffusion process itself, turning into so-called anomalous diffusion. The distinctive trait of standard diffusion is a square root dependence on time of the mean displacement in space,
$<\delta x^2> = D \delta t$. Anomalous diffusion generalizes this relation to a generic exponent $p$, namely $<\delta x^2> = D_p \delta t^p$. Standard diffusion is recovered for $p=1$, while $1<p<2$ corresponds to super-diffusion and $0<p<1$ to sub-diffusion, respectively.

The anomalous diffusion coefficient $p$ encapsulates complex transport phenomena, resulting in both faster (super) and slower (sub) dynamics than standard diffusion. Generally speaking, this is the result of cross-correlations between the moving molecules and their environment; constructive/destructive
correlations promote hyper/hypo-diffusion, respectively. On the one hand, cytoskeleton density hinders the free displacement of the particle, leading to subdiffusion. On the other hand, the cytoskeleton elasticity combined with thermal bending contributes superdiffusion.

We note that in this case the diffusion coefficient as we know it from the random walk theory, the limit of the ratio $\delta x^2/\delta t$ as $\delta t \to 0$, is no longer a well-posed physical quantity. Indeed, this limit returns zero and infinity for hyper and hypo-diffusion, respectively.
The correct limit is instead $D_p = \delta x^2/\delta t^p$, which has no longer the dimensions of a diffusion coefficient, the length squared over time. We see that the anomalous transport, a hallmark of (intra)cellular transport, represents another type of extrinsic effects setting transport properties in life processes.

\section{Sensitivity and fine-tuning}

Fine-tuning of fundamental constants refers to relative little variation of constants above which an essential physical process (e.g., stability of protons and neutrons, stellar processes needed for synthesis of heavy elements, carbon production as a result of the Hoyle resonance and so on) is disabled \cite{barrow,barrow1,carrbook,finebook,carr1,carr,cahnreview,hoganreview,adamsreview,uzanreview,uzan1}. For some processes and relevant fundamental constants, these variations are often between a few per cent to fractions of per cent. This fine tuning originates from our physical models where a property is a fast-varying function of fundamental constants or their combinations so that small changes of fundamental constants imply large property changes \cite{barrow}.

We can now compare how sensitive viscosity is to variations of intrinsic and extrinsic effects. The minimal viscosity in Eq. (\ref{etamin}) and the intrinsic effects are set by

\begin{equation}
\eta_{min}\propto e^6 \hbar^{-5} m_p^{1/2} m_e^{5/2}
\end{equation}

\noindent and

\begin{equation}
\eta(S)\propto\eta_0 (S/S_0)^{-1/3}
\end{equation}

By way of illustration, a factor of $10$ change in the minimal viscosity requires a $10^3$ change in the shear rate and $10^{1/6} \sim 1.5$ in the electron charge, $10^{-1/5} \sim $ of the Planck constant, $10^{2}$ in the proton mass and $10^{2/5} \sim 3$ in the electron mass. Conversely, a small change of say $0.01$ of the electron charge would cause a small $(1+0.01)^6-1 = 0.06$, change of viscosity. The same change corresponds to a change $x$ of the shear rate given by $(1+x)^{-1/3} = 1.06$, $x \sim -0.18$. This is 18 times higher but still a comparatively small change and is within the physiological range of variation of the shear rate.
The condition for extrinsic and intrinsic effects to be comparable is

\begin{equation}
(1+ \epsilon_f)^{\alpha_f} = (1+\epsilon_e)^{\alpha_e}
\end{equation}

\noindent where $\epsilon_{f,e}$ is the relative change of the intrinsic fundamental and
effective couplings, respectively, and $\alpha_{f,e}$ are the corresponding exponents.

If both $\epsilon$'s are well below 1, the above relation simplifies to:

\begin{equation}
\epsilon_e \sim \epsilon_f \frac{\alpha_f}{\alpha_e}
\end{equation}

\noindent showing that the ratio of the changes is dictated by the ratio of the exponents.
This implies that even a large ratio such as $18$ still keeps $\epsilon_e$ sufficiently small to be easily realizable via environmental changes.

As an interim summary, we saw that both extrinsic and intrinsic factors affecting viscosity need to be taken into account when estimating the bio-friendly range of fundamental constants from life processes. We also saw that the relative role of extrinsic and intrinsic factors depends on the range in which these intrinsic and extrinsic factors vary. This range can be different in different life processes involving flow and hence need to be addressed separately.

\section{Viscosity of complex fluids and fundamental constants}

Blood is particularly significant in our discussion because, as compared with water, a complex
but still a molecular fluid, and involves a much higher level of physiological organization: red
blood cells (RBC) are not molecules but highly organized microscale biological structures.
In this respect, blood might be expected to be largely governed by extrinsic effects. In this regard, it is instructive to consider the value of blood viscosity at rest, which we take of the order of 10$^{-3}-10^{-2}$ Pa$\cdot$s \cite{sauroBLOOD}, giving kinematic viscosity of 10$^{-6}-10^{-5}$ m$^2$/s. This is 13-14 orders of magnitude above what might be expected from evaluating kinematic viscosity using the molecular mass \cite{sciadv1} as $\hbar/M_{RBC}\sim 10^{-19}$ m$^2$/s, where $M_{RBC}$ is the mass of the cell, but is close to the fundamental kinematic viscosity $\nu_f$ in Eq. \eqref{num2}. We note that the fundamental kinematic viscosity corresponds to the viscosity minimum, whereas the observed viscosity can be higher depending on temperature and pressure \cite{sciadv1}. This is consistent with the observed blood viscosity being higher than the fundamental viscosity.

The closeness between the observed blood viscosity and theoretical fundamental viscosity is remarkable because it shows that even in a highly complex mesoscale structure such as a red blood cell, which contains about one trillion protons organized across many layers of biological and physiological complexity far above the quantum level, the bare kinematic viscosity still carries a memory of fundamental physical constants, regardless of the large mass of the red blood cell which is clearly a classical macroscopic body from the standpoint of quantum mechanics.
A tentative explanation can be discussed as follows.

The kinematic viscosity is a collective property emerging from underlying molecular interactions, as reflected by the relation:

\begin{equation}
\nu = \lambda v_{th}
\end{equation}

\noindent where $v_{th} = \sqrt{k_{\rm B}T/m}$ is the thermal speed, $\lambda = v_{th} \tau$
is the mean free path (scattering length) and $\tau$ is the mean collision time.






The scattering length $\lambda=\frac{\nu}{v_{th}}$ corresponding to kinematic viscosity 10$^{-6}$ m$^2$/s is on the order of nanometers, where $v_{th} \sim 10^3$ m/s at standard temperature. This is comparable to the mean scattering distance (mean free path) in water. The flow of cells in blood involves interactions operating at cell contacts, whose frequency is governed by the intracellular gap $h$ discussed earlier in this paper. Let us rewrite the intracellular gap in general terms as


\begin{equation}
\frac{h}{2R}=\left(\frac{\phi_{pack}}{\phi}\right)^{\frac{1}{3}}-1 = (1+\xi)^{\frac{1}{3}}-1 \sim \frac{\xi}{3}
\end{equation}

\noindent where $\xi \equiv (\phi_{pack}-\phi)/\phi$, $2R$ is the cell diameter as before and $\phi_{pack}$ is the packing fraction of the blood configurations.

As discussed earlier, $\phi \sim 0.45$ on average, while $\phi_{pack}$ may change depending on the local configuration, as well as on the shape of the RBC's (ellipsoids, discoids). We observe that $\phi$ and $\phi_{pack}$ are close enough to develop fluctuating intracellular gaps of the order of nanometers and comparable to $\lambda$. These ``near-contact'' interactions are well known to play a crucial role in shaping up the mechanical and rheological properties of a large variety of soft materials \cite{sauroNCIJFM,frenchRMP}, including the ones relevant to human body.

The fact that the mesoscale objects, like cells, interact on nanometric scales acting as an effective mean free path in a statistical mechanics description of their transport properties, offers a plausible reason why their kinematic viscosity is affected by intrinsic effects and quantum mechanics in particular: the interactions between contact points are affected by chemical bonding as is the case in simple liquids. As discussed in Section \ref{intrinsic}: this gives rise to the intrinsic viscosity set by fundamental condensed matter properties, the length scale $a_{\rm B}$ in Eq. (\ref{bohr}) and the energy scale $E_{\rm R}$ in Eq. (\ref{rydberg}) (these two properties are essentially quantum-mechanical and do not have a sensible limit $\hbar\rightarrow 0$). Hence, viscosity of a complex fluid with significant extrinsic effects such as blood is nevertheless affected by intrinsic effects and fundamental constants acting at the contact points.

\section{Summary}

We discussed both intrinsic and extrinsic contributions to viscosity which are at operation in life processes.
A direct inspection of the physiological values of blood viscosity at rest reveals a remarkable near-match with the fundamental kinematic viscosity. This implies that, notwithstanding the four decades separation in space and the
many layers of biological and physiological complexity of red blood cells, the kinematic viscosity of blood
still keeps clear memory of the values of fundamental physical constants.
A possible explanation for this remarkable property is that the near-contact interactions
between RBC's occurs at scales comparable with the fundamental De Broglie wavelength.

Our results give the following recipe to calculate the constraints on fundamental constants from life processes. We can identify several most important life processes where viscosity sets the motion central to each process. Let $\eta_d$ be the upper value of viscosity above which a life process is disabled. Mechanisms for such a disabling can vary and include, for example, a transition corresponding to the explosive increase of the coagulation rate in biological fluids such as protein solutions and blood. We can then use the equations discussed here, such as Eq. (\ref{etaext}), to account for the extrinsic effects. This will result in constraints on the intrinsic (bare) viscosity $\eta_0$ in Eq. (\ref{etaext}). The constraints on fundamental constants will then follow from Eqs. such as (\ref{num1}) and (\ref{etamin}) and, more specifically, from accompanying inequalities setting the bio-friendly window for fundamental constants \cite{biofriendly}.

We are grateful to L. Noirez, U. Windberger and A. Zaccone for discussions and EPSRC for support. S. S. research was supported by the ERC-PoC grant Droptrack (Fast and automated droplet tracking tool for microfluidics, contract n. 101081171).


\end{document}